\documentclass[a4paper,fleqn,usenatbib,useAMS]{mnras}

\usepackage{graphicx}	
\usepackage{amsmath}	
\usepackage{amssymb}	
\usepackage{bm}		
\usepackage{pdflscape}	

\newcommand{\ha}{H$\alpha$}
\newcommand{\oiii}{[{\sc Oiii}]}
\newcommand{\hb}{H$\beta$}
\newcommand{\nii}{[{\sc Nii}]}
\newcommand{\oii}{[{\sc Oii}]}

\newcommand{\nbh}{${\rm NB}_H$}
\newcommand{\nbk}{${\rm NB}_K$}

\usepackage[T1]{fontenc}
\usepackage{ae,aecompl}

\usepackage{txfonts}

\title[Comparison between H$\alpha$ and {[OIII]} emitters at $z$=2.23]
{[OIII] emission line as a tracer of star-forming galaxies at high redshifts: 
Comparison between H$\alpha$ and [OIII] emitters at $z$=2.23 in HiZELS}
\author[T. L. Suzuki et al.]{T. L. Suzuki,$^{1,2}$\thanks{E-mail: {suzuki.tomoko@nao.ac.jp}}
T. Kodama,$^{1,2}$ 
D. Sobral,$^{3,4}$ 
A. A. Khostovan,$^{5}$ 
M. Hayashi,$^{2}$ 
\newauthor
R. Shimakawa,$^{1,2}$ 
 Y. Koyama,$^{6}$ 
K.-i. Tadaki,$^{7}$ 
 I. Tanaka,$^{6}$  
Y. Minowa,$^{1,6}$
M. Yamamoto,$^{1,2}$
\newauthor  
I. Smail,$^{8}$
and P. N. Best$^{9}$
\\ 
$^{1}$Department of Astronomical Science, SOKENDAI (The Graduate University for Advanced Studies), Mitaka, Tokyo 181-8588, Japan \\
$^{2}$Optical and infrared Astronomy Division, National Astronomical Observatory of Japan, Mitaka, Tokyo, 181-8588, Japan \\
$^{3}$Department of Physics, Lancaster University, Lancaster, LA1, 4YB, UK  \\
$^{4}$Leiden Observatory, Leiden University, PO Box 9513, NL-2300 RA Leiden, the Netherlands  \\
$^{5}$Department of Physics \& Astronomy, University of California, 900 University Ave., Riverside, CA 92521, USA \\
$^{6}$Subaru Telescope, National Astronomical Observatory of Japan, N. A$\textquoteright$ohoku Pl., Hilo, HI 96720, USA  \\
$^{7}$Max-Planck-Institut f\"{u}r Extraterrestrische Physik, Giessenbachstrasse, D-85748 Garching, Germany \\
$^{8}$Centre for Extragalactic Astronomy, Department of Physics, Durham University, South Road, Durham DH1 3LE, UK \\
$^{9}$SUPA, Institute for Astronomy, Royal Observatory of Edinburgh, Blackford Hill, Edinburgh EH9 3HJ, UK }
\date{Accepted 2016 July 07. Received 2016 June 30; in original form 2016 February 11.}

\pubyear{2016}

\begin{document}
\label{firstpage}
\pagerange{\pageref{firstpage}--\pageref{lastpage}}
\maketitle

\begin{abstract}
We investigate the properties of $z$=2.23 \ha\ and \oiii$\lambda$5007 emitters  
using the narrow-band-selected samples obtained from the High-$z$ Emission Line Survey (HiZELS: \citealt{sobral13}). 
We construct two samples of the \ha\ and \oiii\ emitters 
and compare their integrated physical properties.  
We find that the distribution of stellar masses, dust extinction, star formation rates (SFRs), and specific SFRs, is not statistically different 
between the two samples. 
When we separate the full galaxy sample into three subsamples 
according to the detections of the \ha\ and/or \oiii\ emission lines, 
most of the sources detected with both \ha\ and \oiii\ 
show ${\rm log(sSFR_{UV})}$$\gtrsim$-9.5. 
The comparison of the three subsamples 
suggests that sources with strong \oiii\ line emission tend to have 
the highest star-forming activity out all galaxies that we study. 
We argue that the \oiii\ emission line can be used as a tracer of star-forming galaxies at high redshift, 
and that it is especially useful to investigate star-forming galaxies at $z$$>$3, 
for which \ha\ emission is no longer observable from the ground. 
\end{abstract}


\begingroup
\let\clearpage\relax
\endgroup
\newpage

\section{Introduction} 
\label{intro}

Emission lines from regions ionized by hot, young massive stars 
are useful as indicators of star formation in distant galaxies.   
Imaging observations with narrow-band (NB) filters, 
which can capture redshifted strong emission lines, 
are a powerful method to construct a star-forming galaxy sample at a particular redshift  
(e.g. \citealt{bunker95,malkan96,moorwood00,geach08,sobral09,sobral13,tadaki13,an14,stroe15}). 
The \ha\ emission line is one of the best tracers of star formation  
because it is less affected to dust extinction 
than the ultraviolet (UV) light 
and the relation between star-formation rates (SFRs) and \ha\ luminosities 
has been well calibrated in the local Universe (e.g. \citealt{hopkins03}). 
This seems to hold at higher redshift. 
In addition, \ha\ selection has the advantage of recovering the full population 
of star-forming galaxies \citep{oteo15}.

However, the redshift range for \ha\ selection 
 is limited to $z$$<$2.8   
because \ha\ emission is no longer easily observed beyond $z$$\sim$2.8 with ground-based telescopes. 
Space telescopes, such as the {\it Spitzer},   
have been used for the \ha\ emission line galaxy survey at higher redshift, 
$z$$\gtrsim$4.0, but only using broad band photometry 
(e.g. \citealt{shim11,smit15}), 
which is therefore only sensitive to the highest equivalent width lines. 
In order to investigate star-forming galaxies at $z$$>$2.8 with NB imaging observations,  
it is necessary to use other emission lines at shorter wavelengths than \ha, 
such as \oii$\lambda$3727, \hb, and \oiii$\lambda5007$. 
 With \oii, \hb, and \oiii\ emission lines, 
 we can reach up to $z$$\sim$5.2, 3.7 and 3.6, respectively, from the ground 
 \citep{khostovan15}.  
\oii\ and \hb\ are relatively weak lines and 
so 
it is more difficult to observe them at higher redshift. 
There is also evidence of higher \oiii/\oii\ line ratios at high redshifts, 
and a potential decline in \oii\ equivalent width \citep{khostovan16}. 
On the other hand,  
strong \oiii\ detections that are comparable to \ha\ 
from high-redshift star-forming galaxies have been reported by 
recent near-infrared (NIR) spectroscopic observations 
(e.g. \citealt{holden14,masters14,steidel14,shimakawa15,shapley15}). 
Such strong \oiii\ emission indicates extreme interstellar medium (ISM) conditions in high-redshift galaxies, 
and this is likely to be due to their lower metallicities and/or higher ionization parameters (e.g. \citealt{nakajima14}). 
Also, \oiii\ emission in the rest-frame optical is less sensitive to dust extinction than the UV light. 
In these respects, it is expected that the \oiii\ emission line can be used 
to select star-forming galaxies at $z$$\sim$3--3.6, 
corresponding to  
$\sim$1--1.5 billion years before the peak epoch of galaxy formation and evolution at $z$$\sim$2 
(e.g. \citealt{hopkins06,khostovan15}). 
Some studies have been constructing \oiii\ (+ \hb) emitter samples at $z$$>$3, and 
have investigated their star-forming activity or the evolution of the luminosity function 
\citep{teplitz99,maschietto08,labbe13,smit14,suzuki15, khostovan15,khostovan16}. 
Star-forming activity and other physical properties of galaxies at $z$$>$3 have also been investigated 
with UV-selected galaxies, such as Lyman Break Galaxies 
(e.g. \citealt{stark09,gonzalez10,reddy12,stark13,tasca14}). 
By using \oiii\ emission as a tracer at $z$$>$3, 
it is expected that we can obtain further understanding about galaxy formation and evolution 
before the peak epoch of galaxy assembly.

However, there are possible biases resulting from the use of \oiii\ emission as a star-forming indicator. 
For example,  
the \oiii\ emission line originates from ionized regions not only caused by hot, young massive stars 
in star-forming regions but also by active galactic nuclei (AGNs; e.g. \citealt{zakamska04}). 
Due to the shorter wavelength of \oiii\ (5007\AA) with respect to \ha\ (6563\AA), 
samples of \oiii-selected galaxies may be inherently biased against dusty systems in comparison to 
an \ha-selected sample. 
As already mentioned above, 
galaxies with strong \oiii\ emission might be biased towards 
galaxies with lower metallicities and/or higher ionization states, 
resulting in a potential bias towards galaxies with lower stellar masses 
given the well-known mass--metallicity relation of star-forming galaxies 
(e.g. \citealt{tremonti04,erb06,stott14,troncoso14}).

At $z$$\sim$0 and $z$$\sim$1.5, the selection biases between \ha\ and \oiii\  
have been investigated 
using the SDSS galaxies and FMOS-COSMOS galaxies \citep{silverman14} 
by Juneau et al. (in preparation, private communication).   
Their samples are selected based on \ha\ and \oiii\ luminosities.  
\citet{mehta15} investigated the relation between \ha\ and \oiii\ luminosity 
for galaxies at $z$$\sim$1.5 in the 
{\it HST}/WFC3 Infrared Spectroscopic Parallel Survey 
(WISP; \citealt{atek10}), 
and derived the \ha--\oiii\ bivariate luminosity function at $z$$\sim$1.5.  
\citet{sobral12} and \citet{hayashi13} investigated the relation 
between the NB-selected \ha\ and \oii\ emission line galaxies at $z$$\sim$1.47. 
They discussed the lack of redshift evolution of the \oii/\ha\ ratio from $z$$\sim$0 to 1.5 \citep{sobral12}, 
and also found that the \oii-selected galaxies tend to be biased towards 
less dusty galaxies with respect to the \ha-selected galaxies \citep{hayashi13}. 
At $z$$>$2, 
some studies have already performed the comparisons between the samples of star-forming galaxies 
selected by different optical emission lines, such as \ha\ and Ly$\alpha$, or by other selection methods 
(e.g. \citealt{oteo15,hagen15,matthee16,shimakawa16}).    
However, 
the comparison of the physical quantities 
between \ha\ and \oiii-selected galaxy samples has not been done yet at $z$$>$2. 
Such a comparison is necessary in order to accurately interpret results from \oiii\ surveys at $z$$>$3.

In this study,  
we use the NB-selected \oiii\ and \ha\ emission line galaxies 
at $z$=2.23, obtained by HiZELS 
(the High-$z$ Emission Line Survey; \citealt{best13,sobral09,sobral12,sobral13,sobral14}),  
a large NB imaging survey. 
The \oiii\ and \ha\ emission lines are observed using the \nbh\ and \nbk\ filters, respectively. 
This combination of NB filters allows for the creation of a suitable sample to investigate 
possible selection biases between the \oiii\ and \ha\ emission line galaxies at high redshift. 
 We compare the integrated physical quantities, such as stellar masses, 
dust extinction, and star formation rates (SFRs), 
and investigate whether there are any systematic differences between these physical quantities.   
Some galaxies are detected with both the \oiii\ and \ha\ emission lines.    
We investigate the physical properties of the galaxies depending on the detectability 
of their \oiii\ and \ha\ emission lines.

This paper is organized as follows: 
In \S \ref{data}, we briefly introduce the NB imaging survey, HiZELS, 
and describe how the \oiii\ and \ha\ emission line galaxies at $z$$\sim$2 are selected.  
We also present the method for deriving the integrated physical quantities. 
Then, we show our results in \S \ref{results}. 
We present the relationship between stellar masses and SFRs for the two emitter samples,  
and compare the number distribution of the global physical quantities between the \ha\ and \oiii\ emitters.  
Moreover, we divide our full sample into three subsamples according to the detections 
of the \ha\ and/or \oiii\ emission lines,
and compare the distributions of physical quantities among the three subsamples. 
We summarize this study in \S \ref{summary}. 
We assume the cosmological parameters of $\Omega_{\rm m}=0.3$, 
$\Omega_{\Lambda}=0.7$, and $H_{\rm 0}=70 \ [{\rm km\ s^{-1}Mpc^{-1}}]$.
Throughout this paper all the magnitudes are given 
in AB magnitude system \citep{oke83}, 
and the Salpeter initial mass function (IMF; \citealt{salpeter55}) 
is adopted for the estimation of the stellar masses and SFRs 
\footnote{Stellar mass estimated assuming the Salpeter IMF can be scaled to 
those assuming the Chabrier \citep{chabrier03} and Kroupa \citep{kroupa02} IMF 
by dividing by a factor of $\sim$ 1.7 and 1.6, respectively \citep{pozzetti07,marchesini09}.}.

\section{Data and analysis}
\label{data}
\subsection{NB imaging survey in the COSMOS field}
\label{HiZELS}

HiZELS is a systematic NB imaging survey 
using NB filters in the $J$, $H$, and $K$-bands of the Wide Field CAMera (WFCAM; \citealt{casali07}) 
on the United Kingdom Infrared Telescope (UKIRT), 
and the NB921 filter of the Suprime-Cam \citep{miyazaki02} on the Subaru Telescope \citep{sobral12,sobral13}.  
Emission line galaxy samples used in this study are based on the HiZELS catalogue 
in the Cosmological Evolution Survey (COSMOS; \citealt{scoville07}) field.

\begin{figure}
\begin{center}
\includegraphics[width=0.7\columnwidth, angle=-90]{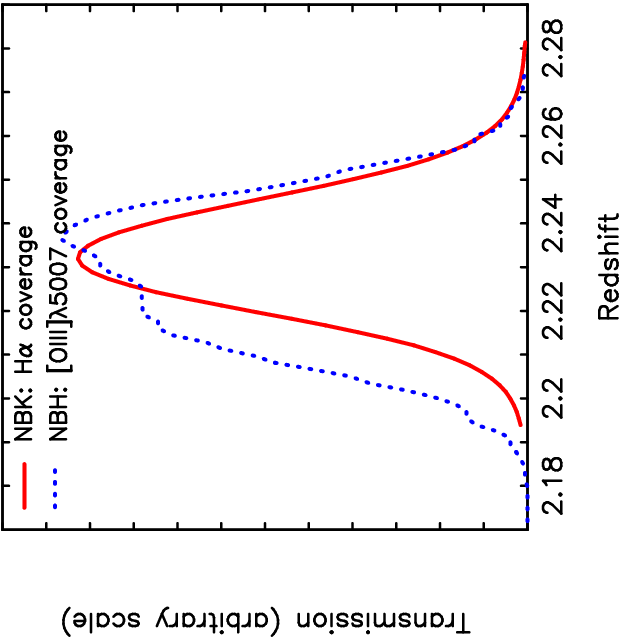}
\caption{A comparison between the ${\rm NB}_H$ redshift coverage for  the \oiii$\lambda$5007 emission line (dotted line)  
and the ${\rm NB}_K$ redshift coverage for the \ha\ emission line (solid line). 
The redshift coverage of \oiii$\lambda$5007 is slightly wider than that of \ha\ as mentioned in \S 2.2.3.  
}
\label{figure1}
\end{center}
\end{figure}

An advantage in the design of the NB filters is the symmetry in respect to their wavelength centers, 
such that an \oiii\ emission can be detected in \nbh\ and \ha\ in \nbk(${\rm H_2 S1}$) with both detections occurring at $z$=2.23 
(Table \ref{sec:filters} and Figure \ref{figure1}). 
\citet{matthee16} has recently presented NB392 observations 
with the Isaac Newton Telescope  
which target Ly$\alpha$ at $z$=2.23 in a similar way, 
and investigated the Ly$\alpha$ properties of the $z$=2.23 \ha\ emitters.

\begin{table}
\begin{center}
\caption{NB filters in the $H$ and $K$-band used in HiZELS \citep{sobral13}.}
\label{sec:filters}
\begin{tabular}{cccl} \hline
Filter & $\lambda_{\rm c}$ [$\mu$m] & FWHM [\AA] & \ \ \ Redshift coverage\\  \hline \hline
\nbh\  & 1.617 & 211 & 2.23 $\pm$ 0.021 \  for \oiii$\lambda5007$ \\
\nbk\  & 2.121 & 210 & 2.23 $\pm$ 0.016 \ for \ha\  \\ \hline
\end{tabular}
\end{center}
\end{table}%

\subsection{Selection of \oiii\ and \ha\ emitters}
\label{selectionmethod}

The catalogs of \ha\ emitters and \oiii\ emitters at $z$=2.23 used in this study 
are taken from \citet{sobral13} and \citet{khostovan15}, respectively. 
The selection criteria of these emitters 
are described in detail in the two papers.  
Here we briefly summarize the selection methods in the following subsections. 
 
\subsubsection{Selection of NB excess sources}
\label{NBemitter}

The sources that are significantly brighter in the NB than in the broad-band (BB) 
are selected as NB emitters 
using BB-NB color versus NB magnitude diagrams (Figure 3 in \citealt{sobral13}). 
A parameter $\Sigma$ is introduced 
to quantify the significance of a NB excess relative to 1$\sigma$ photometric error \citep{bunker95}.  
This parameter $\Sigma$ is represented as a function of NB magnitude as follows \citep{sobral13}:  

\begin{equation}
\Sigma = \frac{1 - 10^{-0.4({\rm BB-NB})}}{10^{-0.4({\rm ZP-NB})} \sqrt{\pi r^2_{\rm ap} (\sigma ^2_{\rm NB} + \sigma ^2_{\rm BB})}}, 
\end{equation}

\noindent
where NB and BB are NB and BB magnitudes, 
ZP is the zero-point of the NB (the BB images are scaled to have the same ZP as the NB images), 
$r_{\rm ap}$ is the aperture radius in pixel, 
and $\sigma_{\rm NB}$ and $\sigma _{\rm BB}$ are the rms 
per pixel of the NB and BB images, respectively \citep{sobral13}. 
The criterion is set to be $\Sigma>3$ to sample secure NB emitters.  
A rest-frame equivalent width (EW) limit of ${\rm EW_{rest}}=25 \AA$ is also applied \citep{sobral13,khostovan15}.   
These selection criteria are applied for both \nbh and \nbk.

\subsubsection{Redshift identification}
\label{redshiftseparation}
The redshift identification of the NB emitters for \nbh\ and \nbk\ is performed based on the photometric redshifts, 
broad-band colors (color--color selections) and the spectroscopic redshifts.   
Here we give the priorities in the following order (from higher to lower):  
(1) spectroscopic redshifts, (2) photometric redshifts, and (3) color--color selections 
\citep{sobral13,khostovan15}. 
If the sources are spectroscopically confirmed to be the targeted line emitters, 
they are firmly identified as the \ha\ or \oiii\ emitters. 
The numbers of such \ha\ and \oiii\ emitters which are confirmed with the spectroscopic redshifts 
are however only 
three and one, respectively \citep{sobral13,khostovan15}. 
Secondly, if the sources have the photometric redshifts within $1.7<z_{\rm phot}<2.8$, 
they are robustly identified as the \ha\ or \oiii\ emitters at $z$$\sim$2.23. 
Here, the photometric redshifts are taken from the catalog of \citet{ilbert09}.

Color--color diagrams are also applied for the redshift separation of the emitters. 
For the \nbk\ emitters, 
the ($z-K$) versus ($B-z$) color--color diagram is used to sample additional 
faint \ha\ emitters at $z$$\sim$2, which lack reliable photometric redshifts.  
In addition to the $BzK$ selection, 
the photometric redshift criteria of $z_{\rm photo}<3.0$ 
or the color--color diagram of ($B-R$) versus ($U-B$) are applied 
to remove higher-redshift sources \citep{sobral13}. 
For the \nbh\ emitters, 
the $BzK$ color--color diagram is used to remove the foreground at $z$$<$1.5, 
and additionally, the ($z-K$) vs ($i-z$) diagram is applied to separate the potential $z$=1.47 \ha\ emitters. 
In order to remove the higher redshift sources, 
the ($V-z$) versus ($U-V$) diagram is used \citep{khostovan15}.

Note that \citet{sobral13} and \citet{khostovan15} 
applied slightly different color--color diagrams 
because 
the other strong emission lines that could contaminate the samples 
(and hence the redshifts of the foreground and background galaxies 
that need to be excluded) are different for the two NB filters.     
We confirm that 
there is no systematic difference in the distributions of the \ha\ and \oiii\ emitters at $z$$\sim$2.23 
on any of the color--color diagrams mentioned above. 
We consider that the color--color selection for the \nbk\ and \nbh\ emitters 
are consistent with each other.  
In this study, we follow the color--color selection criteria for each NB filter.  
This difference does not cause any systematic differences between the two emitter samples.

The number of the redshift-identified \ha\ and \oiii\ emitters at $z$=2.23 
is 513 and 172, respectively.

\subsubsection{AGN contribution of the two emitter samples}
\label{IRACcolor}

We use 
the X-ray observations 
and {\it Spizer}/IRAC colors 
to investigate the contribution of AGNs to our \ha\ and \oiii\ emitter samples at $z$=2.23. 

The redshift-identified \ha\ and \oiii\ emitters are matched with the X-ray point source catalog  
from the {\it Chandra} COSMOS Legacy survey \citep{civano16} 
in order to identify obvious AGNs. 
The fraction of the X-ray-detected sources is only 
2.3\% and 3.5\% for the \ha\ and \oiii\ emitters, respectively.  
We remove these X-ray-detected sources from the two emitter samples.

We also estimate the fractions of obscured AGN candidates in the \ha\ and \oiii\ emitters. 
The colors in {\it Spitzer}/IRAC four channels are commonly used 
to identify obscured AGNs (e.g. \citealt{lacy07,stern05,donley08}). 
We here use only the sources detected at more than 
the $2\sigma$ level in all four channels of IRAC, 
which limited our \ha\ and \oiii\ emitter samples to only 15\% and 17\% (76 and 29 sources), respectively. 
When we use the $S_{\rm 5.8}/S_{\rm 3.6}$--$S_{\rm 8.0}/S_{\rm 4.5}$ diagram with 
the selection criteria of \citet{donley12}, 
the fraction of the emitters which can be classified as AGNs  
is $\sim$14\% for both the \ha\ and \oiii\ emitter samples. 
This fraction must be an over estimation of the true fraction. 
Given the fact that the bright \ha\ emitters are more likely to be AGNs \citep{sobral16}, 
using only the emitters that are bright enough in all four IRAC channels 
might cause a higher AGN fraction than the true fraction.

We note that the fractions of the X-ray-detected sources or IRAC-color-selected AGNs are not different 
between the \ha\ emitters and \oiii\ emitters at $z$=2.23, 
indicating that the \oiii\ emitters do not necessarily show the higher fraction of AGNs 
as compared to the \ha\ emitters.

\subsubsection{Final samples of \oiii\ and \ha\ emitters at $z$=2.23}
\label{finalsample}

The HiZELS NB survey in the COSMOS field covers a very wide area of 1.6 ${\rm deg^2}$,  
but the survey depth is different among the WFCAM pointings and among different NB filters (see \citealt{sobral13}). 
In this study, 
in order to ensure the same flux limit for both NB-selected samples, 
we use the sources in the deepest pointings only.  
The minimum exposure times are 
107 ks and 62.5 ks for \nbh\ and \nbk, respectively. 
The survey area is then limited to the central $\sim$ $0.2$ ${\rm deg^2}$. 
The 3$\sigma$ limiting fluxes for \nbh\ and \nbk\ 
 are $3.60\times10^{-17}\ {\rm erg\ s^{-1}cm^{-2}}$ 
 and $2.96\times10^{-17}\ {\rm erg\ s^{-1}cm^{-2}}$, respectively.  
In this study, 
we apply the same flux limit of $3.6\times10^{-17}\ {\rm erg\ s^{-1}cm^{-2}}$ 
(\oiii\ flux for \nbh\ and \ha +\nii\ flux for \nbk). 
As a result, 49 \oiii\ emitters and 44 \ha\ emitters remain in our final samples.

In \S \ref{dual}, 
we focus on the galaxies detected with both \oiii\ and \ha\ emission lines 
(hereafter, dual emitters). 
When we search for the counterpart line, 
we lower the line detection threshold, 
as we can trust more the existence of a line at the expected wavelength 
in the other NB filter. 
We thus use the lower \nbh\ (\nbk) excess criteria, namely,  
EW $\gtrsim$15 \AA\ and/or $\Sigma$$\gtrsim$2.  
We find 23 dual emitters in total. 
Among them, ten sources satisfy the original NB excess criteria of both \nbh\ and \nbk, 
indicating that they have the strong \oiii\ and \ha\ emission lines.  
Eleven (two) sources are the \ha\ (\oiii) emitters with the relatively weak \oiii\ (\ha) emission lines.  
The number of sources in each sample used in this study is summarized in Table \ref{sec:numbers}.

 We note that the transmission curves of \nbh\ and \nbk\ filters are not completely matched (Figure \ref{figure1}).   
The wavelength coverage of the two filters are transformed to the redshift space for each line
in Figure \ref{figure1} and Table \ref{sec:filters}. 
It turns out that $\sim$10\% of the \nbh\ redshift coverage (\oiii$\lambda$5007)
is completely out of the \nbk\ redshift coverage (\ha). 
In terms of FWHM ranges of the two NB filters, 
however, 24\% of the \nbh\ coverage is out of the \nbk\ coverage in redshift space. 
This mismatch of the redshift coverage is not critical when we compare the \ha\ and \oiii\ emitters in \S \ref{twosample}. 
However, 
when we consider the sample of galaxies detected only by the \oiii\ emission line,  
the impact of this difference could be larger.  
The redshift mismatch might cause the loss of \ha\ flux 
from the galaxies which actually have the strong enough \ha\ emission line, 
and thus  
the observed \oiii/\ha\ ratio would become different from the intrinsic ratio. 
In such a case, 
the \ha\ flux would seem much lower with respect to \oiii, 
and in the extreme case, the emitter would appear as 
an \oiii\ emitter with no \ha\ emitter counterpart. 
Therefore, 
``\oiii-single-emitters'' (see Table \ref{sec:numbers} for the definition) 
may include some dual emitters, 
and one should use caution when comparing the properties of this subsample with others.

\begin{table*}
\begin{center}
\caption{Summary of samples and subsamples used in this study.}
\label{sec:numbers} 
\begin{tabular}{ccc} \hline
Name & Number of sources & Notes \\ \hline \hline
\ha\ emitters & 44 & -- \\
\oiii\ emitters & 49 & -- \\ \hline {\smallskip}
\ha-single-emitters & 23 & \ha\ emitters with no \oiii\ emitter counterpart \\ {\smallskip}
\oiii-single-emitters & 37 $^{\rm i)}$ & \oiii\ emitters with no \ha\ emitter counterpart \\ 
Dual emitters & 23 & \ha\ + \oiii\ emitters (dual emitters) \\ \hline
\end{tabular}
\end{center}
{
i) We note that the number of the \oiii\ emitters with no \ha\ emitter counterpart is larger than 
that of the \ha\ emitters with no \oiii\ emitter counterpart. 
Some \oiii\ emitters might be \oiii$\lambda$4959 or \hb\ emitters as discussed in \S \ref{NBHcontamination}, 
and thus they do not have \ha\ emitter counterparts. \\
}
\end{table*}%

\subsection{Contribution of \hb\ and \oiii$\lambda$4959 emitters}
\label{NBHcontamination}

There are the two lines which are close to our target line (\oiii$\lambda$5007), 
namely, \oiii$\lambda$4959 and \hb\ at $\lambda$=4861\AA. 
The wavelength coverage of \nbh\ is too narrow 
to include both the \hb\ and \oiii\ lines simultaneously. 
On the other hand, the \oiii\ doublet lines can be detected at the same time at the 
 the opposite edges of \nbh\  for some galaxies in a narrow redshift range of $\Delta z$$\sim$0.01. 
 However, the fraction of such emitters is expected to be small 
 ($\sim$7\% of the \oiii+\hb\ emitters at $z$=1.47; 
 see the full analysis from spectroscopy in \citealt{sobral15}).

However, as is also noted in \citet{khostovan15}, 
\hb\ or \oiii$\lambda$4959 at slightly different redshifts can not be actually distinguished 
from our target \oiii$\lambda$5007 emitters at $z$=2.23 by photometric redshifts and broad-band color selections 
because their redshifts are too close to separate only with photometric data. 
We estimate the fraction of such \hb\ and \oiii$\lambda$4959 emitters included in the \nbh\ emitters 
by considering the luminosity functions of \hb\ and \oiii$\lambda$4959.

The luminosity functions of the \hb\ and \oiii$\lambda$4959 line are determined by converting
the luminosity functions of the \ha\ \citep{sobral13} and \oiii\ line \citep{colbert13} 
assuming the line ratios of \ha/\hb\ $\sim$ 5 incorporating the effects of dust  
based on the observations of \citet{hayashi11} and \citet{shimakawa15},   
and \oiii$\lambda5007$/\oiii$\lambda4959$ = 3 (the theoretical ratio; \citealt{storey00}), respectively.    
With our 3$\sigma$ detection limit of $3.60\times10^{-17}\ {\rm erg\ s^{-1}cm^{-2}}$,  
the contribution of the \hb\ emitters is estimated to be $\sim$3\%.  
Similarly, the contribution of the \oiii$\lambda$4959 emitters is estimated to be 25\%.  
\citet{sobral15} performed the spectroscopic follow-up observations 
for the \oiii\ + \hb\ emitters at $z$$\sim$1.47 obtained by HiZELS. 
They found that the \oiii$\lambda$5007 emitters are dominant population, 
constituting $\sim$50\% of the full sample of the \oiii\ + \hb\ emitters. 
The \oiii$\lambda$4959 and \hb\ emitters represent 27\% and 16\% of the sample, respectively.  
The remaining 7\% of sources are detected with both of the \oiii\ doublet lines at the opposite edges of the NB filter. 
The fraction of \oiii$\lambda$4959 emitters is consistent with our estimation using the luminosity functions, 
although the fraction of the \hb\ emitters of our estimation is smaller than that of \citet{sobral15}.   
This is probably due to the difference of the targeted redshifts ($z$=1.47 versus 2.23). 
Considering that the \oiii\ emission line becomes more prominent at higher redshifts (e.g. \citealt{faisst16}),   
and that the fraction of the \hb\ emitters among the \oiii+\hb\ emitters decreases with luminosity \citep{sobral15}, 
the fraction of the \hb\ emitters might decrease at higher redshifts.

We note that the fraction of \oiii$\lambda$4959 is not negligible. 
However, the galaxies with strong \oiii$\lambda$4959 
should have even stronger \oiii$\lambda$5007 emission given the line ratio.
Therefore, the \oiii$\lambda$4959 emitters  
 and the emitters which are detected with both the \oiii\ doublet lines 
 can also be regarded as the \oiii$\lambda$5007 emitters, 
although they are still missed out from the dual emitter sample. 
Spectroscopic follow-up observations are necessary to accurately quantify the contribution
of the \hb\ and \oiii$\lambda$4959 emitters at $z$$\sim$2.

We note that the grism spectroscopy of  {\it HST} also suffers from blending of 
the \hb, \oiii\ 4959 and 5007 emission lines. 
The wavelength resolution of {\it HST}/WFC3 grism is not enough to resolve the \oiii\ doublet lines. 
In the WISP survey, for example, 
they tried to deblend the \oiii\ 4959 and 5007 lines by using a multi-Gaussian model 
with the fixed \oiii$\lambda$5007/\oiii$\lambda$4959 ratio \citep{atek10,atek14}. 

In summary, 
taking into account the relative strength of \oiii$\lambda$5007 as compared to \oiii$\lambda$4959 and \hb, 
we consider that the majority of the \nbh\ emitters are the \oiii$\lambda$5007 emitters
once we apply the redshift identification in \S \ref{redshiftseparation}. 
Moreover, our \oiii$\lambda$5007 line flux is not contaminated by \oiii$\lambda$4959 or \hb\ at the same redshift for most of the emitters.

\subsection{Estimation of physical quantities}

\subsubsection{Stellar masses}
\label{mass}

Stellar masses are estimated from spectral energy distribution (SED) fitting with
a stellar population synthesis model based on the public code EAZY \citep{brammer08}
and FAST \citep{kriek09}.   
We use 16 photometric band data, 
$FUV$, $NUV$, $u$, $B$, $V$, $g$, $r$, $i$, $I_{\rm c}$, $z$, $J$, $K$, 3.6, 4.5, 5.8, and 8.0 $\mu$m 
obtained from the photometric catalog of \citet{ilbert09}.  
For the sources detected with \nbk, 
\ha+\nii\ line fluxes are subtracted from the $K$-band fluxes before performing the SED fitting. 
The contribution of \ha+\nii\ fluxes corresponds to $\sim$20\% of the $K$-band fluxes 
on average. 
The redshift of the galaxies is fixed to $z$=2.23.    
We use the stellar population synthesis model of \citet{bc03} with the Salpeter IMF \citep{salpeter55}
and the dust extinction law of \citet{calzetti00}. 
We assume exponentially declining star formation history (SFH) in the form of SFR $\sim {\rm exp}(-t/\tau)$, 
with ${\rm log(\tau/yr)}$= 8.5--10.0 in steps of 0.1, 
and the metallicities of 0.004, 0.008, 0.02 (solar) and 0.05, 
similar to the analysis of \citet{sobral14}.

\subsubsection{Star formation rates and dust extinctions}
\label{SFR}

We estimate the rest-frame 1600\AA\ luminosity at $z$=2.23 
using the {\it V}-band photometry from \citet{ilbert09}, 
and convert $L(1600 \AA)$ to the UV star formation rates (SFRs) as mentioned below. 
Dust extinction is corrected for using the slope of the rest-frame
UV continuum spectrum (e.g. \citealt{meurer99,heinis13}). 
The UV slope $\beta$ is defined as $f_{\lambda}$$\propto$$\lambda^{\beta}$. 
We estimate $\beta$ by fitting a linear function 
to the five broad-band photometries from the $B$ to $i$ -bands. 
The slope $\beta$ is converted to the dust extinction $A_{\rm FUV}$ with the following equation from \citet{heinis13}:

\begin{equation}
A_{\rm FUV} = 3.4 + 1.6 \beta.
\end{equation}

\noindent 
Then, the intrinsic flux density $f_{\nu, {\rm int}}$ is obtained from 

\begin{equation}
f_{\nu, {\rm int}} = f_{\nu, {\rm obs}}\ 10^{0.4A_{\rm FUV}}, 
\end{equation}

\noindent
and SFRs are estimated from UV luminosities adopting the equation from \citet{madau98}: 

\begin{eqnarray}\nonumber
{\rm SFR \ (M_\odot yr^{-1})} &=& \frac{4\pi D_L^2  f_{\nu, {\rm int}}}{(1+z) \times 8 \times 10^{27} \ ({\rm erg \ s^{-1} cm^{-2}  Hz^{-1}})} \\
&=& \frac{L({1600} {\rm \AA})}{8 \times 10^{27} \ ({\rm erg \ s^{-1} Hz^{-1}})}, 
\label{eq4}
\end{eqnarray}

\noindent
where $D_L$ is the luminosity distance.

For the sources with \ha\ detections, 
we also derive ${\rm SFR_{H\alpha}}$. 
Since we only obtain \ha+\nii\ fluxes from the \nbk\ imaging observations, 
\nii\ line fluxes should be removed from the total NB fluxes. 
The \nii\ fluxes are removed using 
the correlation of the two line ratio with the EW of \ha+\nii\ as shown in \citet{sobral13}. 
We then estimate ${\rm SFR_{H\alpha}}$ 
using the relation between SFRs and \ha\ luminosities of \citet{kennicutt98}: 

\begin{equation}
{\rm SFR}_{\rm H\alpha} (M_\odot {\rm yr^{-1}}) = 7.9 \times 10^{-42} \frac{L_{\rm H\alpha}}{\rm erg s^{-1}}.  
\end{equation}

\begin{figure}
\begin{center}
\includegraphics[width=0.38\textwidth, angle=-90]{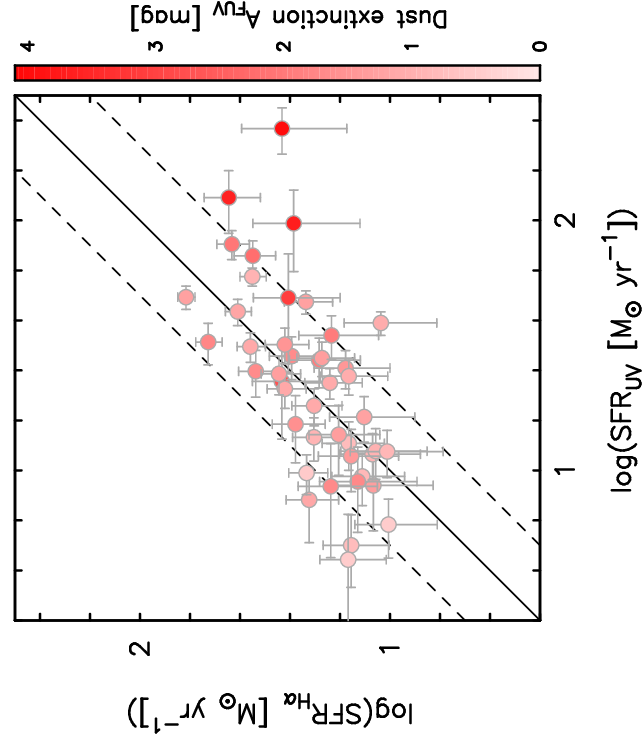}
\end{center}
\caption{
Comparison of SFRs derived from UV luminosities and \ha\ luminosities for the \ha\ emitters at $z$=2.23. 
Dust extinction correction is based on the UV slope $\beta$. 
We assume that there is no extra extinction to emission lines as compared to  the extinction to the stellar continuum. 
The strength of colors are corresponding to the dust extinction $A_{\rm FUV}$. 
The solid line represents the case where the two SFRs are identical, 
and the dashed lines represents the case where a difference between the two is a factor of 2. 
}
\label{compareSFR}
\end{figure}

\noindent
The dust extinction for \ha\ emission    
is estimated from $A_{\rm FUV}$ by assuming the Calzetti extinction law \citep{calzetti00}. 
We assume that there is no extra extinction for the nebular emissions compared to the stellar extinction, 
i.e. $E(B-V)_{\rm stellar}=E(B-V)_{\rm nebula}$ (e.g. \citealt{erb06b,reddy10,reddy15})  
 although this is still under the debate.  
We compare the SFRs derived from two different indicators, UV luminosities and \ha\ luminosities, 
in Figure \ref{compareSFR}. 
 We find that the ${\rm SFR_{UV}}$ and ${\rm SFR_{H\alpha}}$ are broadly consistent with each other 
 within a factor of 2,  
 indicating that we can estimate SFRs reasonably well 
 with UV luminosities with dust correction based on the UV slope.  
For the three sources with the highest ${\rm SFR_{UV}}$, 
their SFRs derived from \ha\ are apparently smaller than those from UV luminosities. 
The difference between the dust extinction to the ionized gas and stellar continuum 
might be involved in such an inconsistency, 
because previous studies targeting the galaxies 
with relatively higher SFRs or higher stellar masses on average 
have indicated that the ionized gas are more strongly affected by dust extinction 
than the stellar continuum 
(e.g. \citealt{forsterschreiber09,kashino13}). 
In the following analysis, 
SFRs of all the sources are represented by ${\rm SFR_{UV}}$. 
This allows us to compare SFRs 
between the \ha\ emitters and \oiii\ emitters with the same SFR indicator.

\section{Results and discussion} 
\label{results}

\subsection{[OIII]/H$\alpha$ flux ratios} 
\label{compareflux}

\begin{figure}
\centering\includegraphics[width=0.4\textwidth,angle=-90]{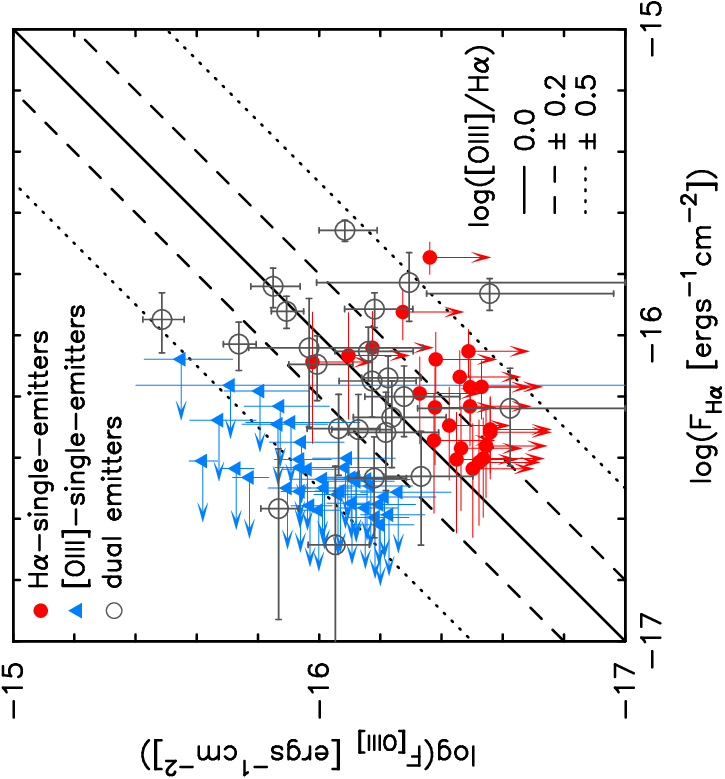}
\caption{
The relation between \ha\ and \oiii$\lambda$5007 flux for the three subsamples.    
Flux of each emission is corrected for dust extinction based on $A_{\rm FUV}$ 
by assuming the extinction curve of \citet{calzetti00} and $E(B-V)_{\rm stellar}=E(B-V)_{\rm nebula}$ 
as mentioned in \S \ref{SFR}. 
Filled circles, filled triangles, and open circles represent the \ha-single-emitters, 
\oiii-single-emitters, and dual emitters, respectively. 
For the \ha-single-emitters and \oiii-single-emitters, 
their \oiii\ or \ha\ fluxes are shown as the upper limit values using the 2$\sigma$ limiting flux.  
The solid, dashed, and dotted line corresponds to log(\oiii/\ha) = 0.0, $\pm$0.2, and $\pm$0.5, respectively.  
}
\label{fluxratio}
\end{figure}

We firstly investigate the \oiii/\ha\ ratios of the samples used in this study. 
In Figure \ref{fluxratio}, 
we compare the dust-extinction-corrected \ha\ and \oiii\ fluxes of 
the three subsamples (Table \ref{sec:numbers}), 
and examine their \oiii/\ha\ ratios.  
For the \ha-single-emitters and \oiii-single-emitters, 
their \oiii\ or \ha\ fluxes are shown as the upper limit values.  
We note that there is an uncertainty of the \oiii/\ha\ ratios 
due to the slightly different wavelength coverage between \nbh\ and \nbk (\S \ref{finalsample}).

\citet{steidel14} and \citet{shapley15} investigated the \oiii/\hb\ ratios 
of star-forming galaxies at $z$$\sim$2. 
Converting their \oiii/\hb\ ratios to \oiii/\ha\ ratios 
by assuming the intrinsic \ha/\hb\ ratio of 2.86 (case B),  
the intrinsic \oiii/\ha\ ratios (log(\oiii/\ha)) are 
estimated to be $\sim$$-0.45$ -- $0.4$ for the samples of \citet{steidel14} 
and $\sim$$-0.45$ -- $0.19$ of \citet{shapley15}. 
In \citet{faisst16}, 
the intrinsic \oiii/\ha\ ratios of star-forming galaxies at $z$$\sim$2.2 are estimated to be roughly $-0.4$ -- $0.2$. 
Figure \ref{fluxratio} shows that our samples have log(\oiii/\ha) $\sim$$-0.6$ -- $0.6$, 
and therefore, although there are still uncertainties regarding the \oiii/\ha\ ratios shown in Figure \ref{fluxratio}, 
our samples of the NB-selected galaxies at $z$=2.23 seem to have 
broadly consistent \oiii/\ha\ ratios with respect to other galaxy samples at the same epoch 
selected with different methods.

\subsection{Stellar mass -- SFR relation}
\label{MS}

\begin{figure}
\begin{center}
\includegraphics[width=0.4\textwidth, angle=-90]{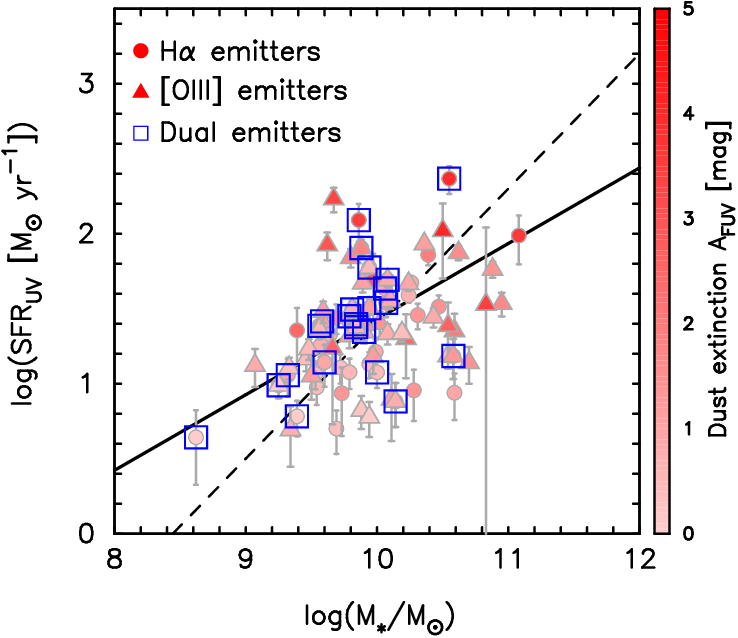}
\end{center}
\caption{
Stellar mass and SFR relation for the NB-selected galaxies at $z$=2.23. 
SFRs are derived from UV luminosities and corrected for dust extinction using the UV slope $\beta$. 
Circles and triangles represent \ha\ emitters and \oiii\ emitters, respectively. 
The strength of the colors corresponds to $A_{\rm FUV}$. 
The dual emitters, as defined in \S \ref{finalsample}, are shown as blue open squares. 
The dashed line shows the stellar mass--${\rm SFR_{UV}}$ relation for the UV-selected galaxies at $z$$\sim$2 from \citet{daddi07}. 
The solid line represents the relation derived from the \ha\ emitters at $z$$=$2.2 with
$M_* \ge 10^{10} {\rm M_{\odot}}$ from HiZELS \citep{koyama13}, although SFRs are based on \ha\ luminosities rather than UV luminosities.
}
\label{MassSFR}
\end{figure}

We investigate the relation between stellar masses and UV-derived SFRs (${\rm SFR_{UV}}$) 
for the NB-selected galaxies at $z$=2.23 (Figure \ref{MassSFR}).  
It is well known that there is an apparent correlation between the stellar mass and SFR 
of moderately star-forming galaxies, 
called the ``star-forming main sequence'' 
(e.g. \citealt{daddi07,elbaz07,noeske07,koyama13,kashino13,whitaker14}).

In Figure \ref{MassSFR}, 
we plot such a diagram using ${\rm SFR_{UV}}$ 
and see a positive correlation for
both of our \ha\ emitters and \oiii\ emitters. 
Importantly, the distributions of the two samples on the stellar mass--SFR diagram 
are not significantly different from each other. 
The dual emitters are also shown in the plot,
and we discuss this population in \S \ref{dual}. 
Each symbol is color-coded according to the dust extinction $A_{\rm FUV}$. 
We can see the trend that the galaxies with higher stellar mass and/or higher SFR 
typically show higher dust extinction.

\subsection{Comparison of physical quantities between the \ha\ and \oiii\ emitters} 
\label{twosample}

\begin{figure*}
\begin{minipage}[cbt]{0.245\textwidth}
\centering\includegraphics[width=1.01\columnwidth,angle=-90]{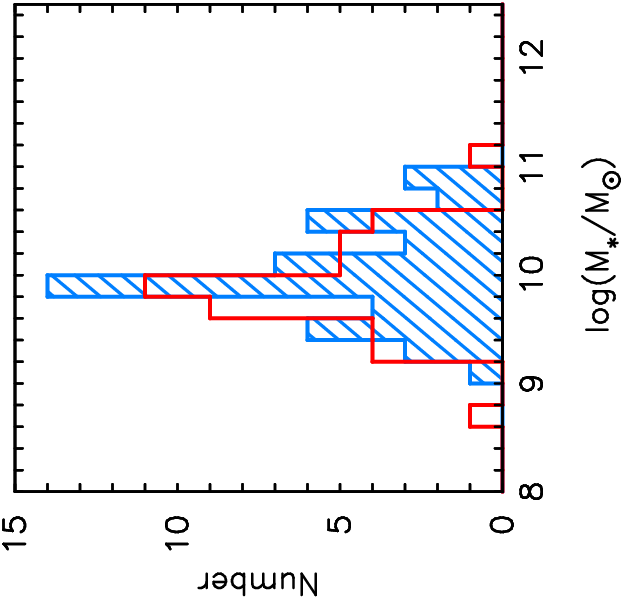}
\end{minipage}
\begin{minipage}[cbt]{0.245\textwidth}
\centering\includegraphics[width=1.01\columnwidth,angle=-90]{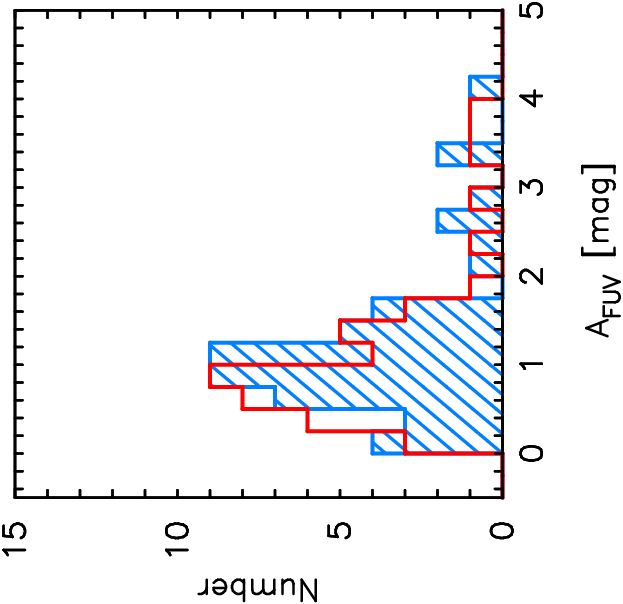}
\end{minipage}
\begin{minipage}[cbt]{0.245\textwidth}
\centering\includegraphics[width=1.01\columnwidth,angle=-90]{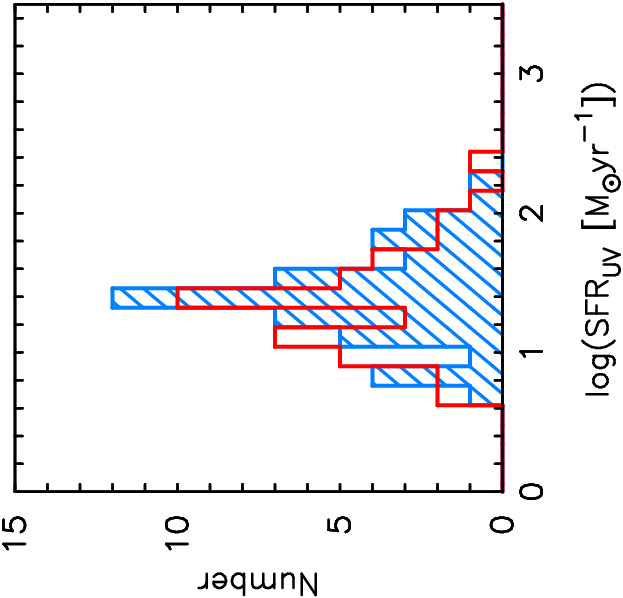}
\end{minipage}
\begin{minipage}[cbt]{0.245\textwidth}
\centering\includegraphics[width=1.01\columnwidth,angle=-90]{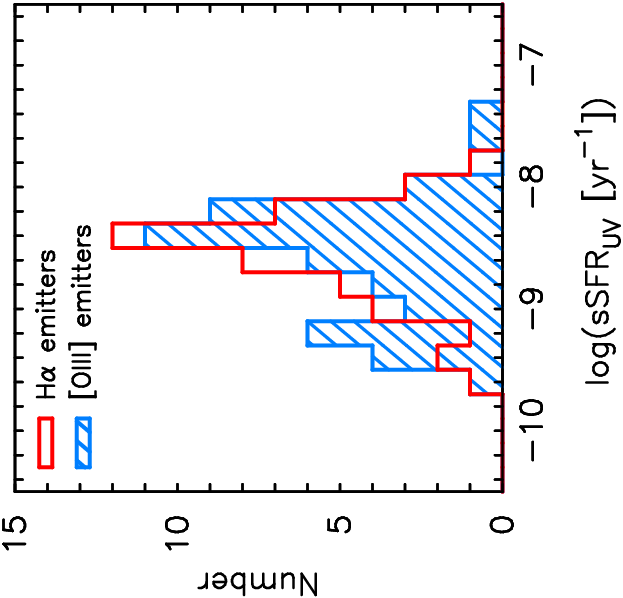}
\end{minipage}
\caption{Number distributions of a  
stellar mass, dust extinction (${A_{\rm FUV}}$), ${\rm SFR_{UV}}$ and ${\rm sSFR_{UV}}$ from left to right. 
The open and hatched histograms represent 
\ha\ emitters and \oiii\ emitters, respectively.  
The $p$-values obtained from the KS-test are listed in Table \ref{pvalue1}. 
}
\label{numdist}
\end{figure*}

In this section, 
we compare the two samples, the \ha\ and \oiii\ emitters at $z$=2.23, 
regardless of whether they are the dual emitters or not. 
We compare the distribution of the integrated physical quantities, such as
a stellar mass, dust extinction, ${\rm SFR_{UV}}$, and specific ${\rm SFR_{UV}}$ (sSFR = ${\rm SFR/M_*\ [yr^{-1}]}$), 
in Figure \ref{numdist}.

In order to investigate whether there are any systematic differences between the two samples, 
we use a Kolmogorov-Smirnov (KS) test. 
The $p$-values from the KS-test are summarized in Table \ref{pvalue1}. 
The $p$-values for all the physical quantities are larger than 0.05, 
meaning that the \ha\ emitters and \oiii\ emitters are consistent with being drawn from
the same population.  
As shown in Figure \ref{numdist}, 
the galaxies selected by \oiii\ emission at $z$=2.23 occupy  
almost the same ranges in their integrated properties  
with those of the galaxies selected by \ha. 
This suggests that 
the \oiii-selected galaxies trace the same general population as 
the \ha-selected galaxies.

\subsection{\ha\ + \oiii\ emitters at $z \sim$ 2.23}
\label{dual}

 \begin{figure*}
\begin{minipage}[cbt]{0.245\textwidth}
\centering\includegraphics[width=1.01\columnwidth,angle=-90]{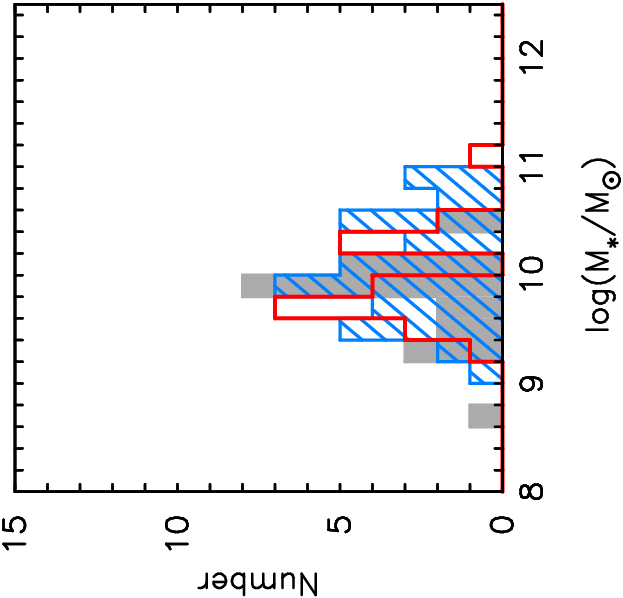}
\end{minipage}
\begin{minipage}[cbt]{0.245\textwidth}
\centering\includegraphics[width=1.01\columnwidth,angle=-90]{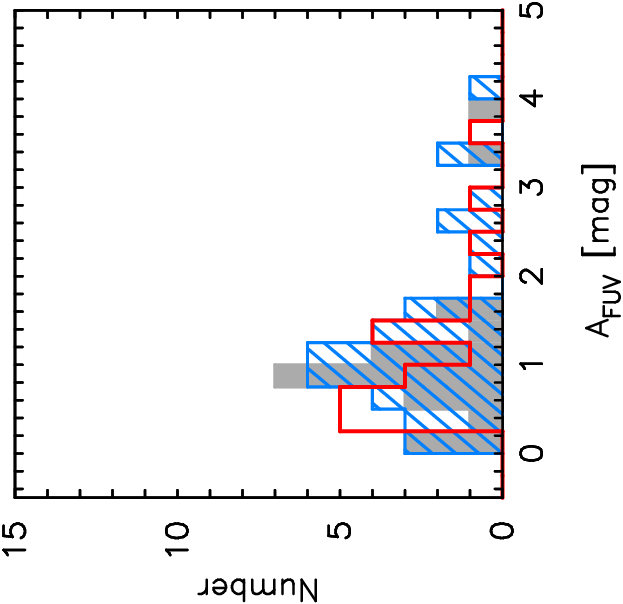}
\end{minipage}
\begin{minipage}[cbt]{0.245\textwidth}
\centering\includegraphics[width=1.01\columnwidth,angle=-90]{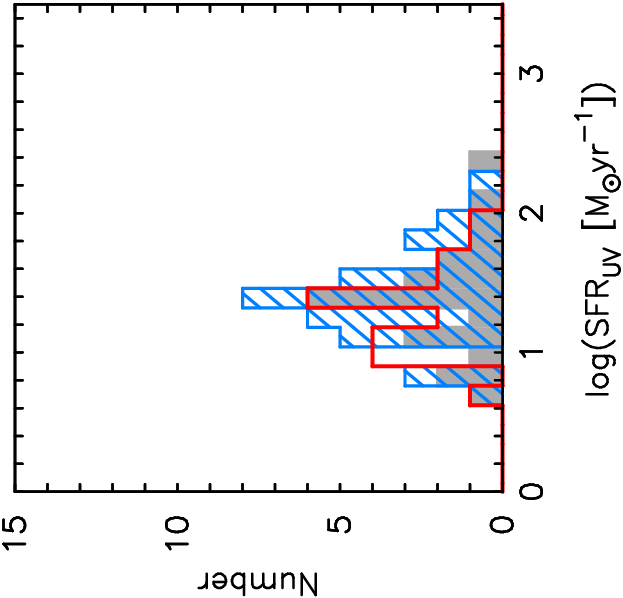}
\end{minipage}
\begin{minipage}[cbt]{0.245\textwidth}
\centering\includegraphics[width=1.01\columnwidth,angle=-90]{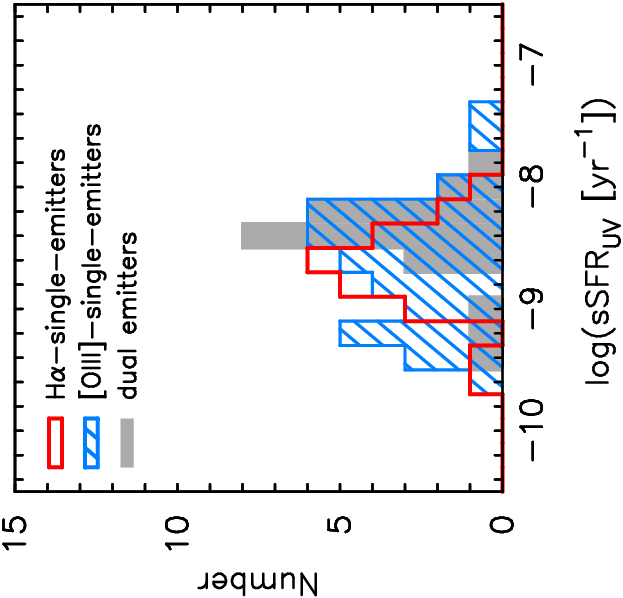}
\end{minipage}
\caption{Number distributions of a 
stellar mass, dust extinction (${A_{\rm FUV}}$), ${\rm SFR_{UV}}$ and ${\rm sSFR_{UV}}$ from left to right. 
The emitter samples are divided into three subsamples, 
 \ha-single-emitters (open histograms), \oiii-single-emitters (hatched histograms), 
and dual emitters (filled histograms). 
The $p$-values are listed in Table \ref{pvalue1}. 
}
\label{numdist3}
\end{figure*}

 \begin{table*}
\begin{center}
\caption{The $p$-values from the KS-test of comparisons of physical quantity distributions
among different emitter samples as shown in Figure \ref{numdist} and Figure \ref{numdist3}.}
\label{pvalue1}
\begin{tabular}{cl|cccc} \hline
  & ${\rm M_*}$ & ${A_{\rm FUV}}$ & ${\rm SFR_{UV}}$ & ${\rm sSFR_{UV}}$ \\ \hline \hline {\smallskip}
  \ha\ emitters versus \oiii\ emitters & 0.76 & 0.69 & 0.56 & 0.77 \\ 
 \ha-single-emitters versus \oiii-single-emitters & 0.85 & 0.63 &  0.56 &  0.57 \\ 
  \ha-single-emitters versus dual emitters & 0.36 & 0.84 & 0.59 & 0.006 \\
  \oiii-single-emitters versus dual emitters & 0.12 & 0.36 & 0.97 & 0.04  \\ \hline 
\end{tabular}
\end{center}
\end{table*}%

\subsubsection{Comparison of the three subsamples}
\label{threesample}

We divide the whole emitter sample into three subsamples according to the
detections of the \ha\ and \oiii\ emission lines as defined in \S \ref{finalsample}, namely,
the galaxies detected with only the \ha\ emission line (\ha-single-emitters), 
the galaxies detected with only the \oiii\ emission line (\oiii-single-emitters), 
and the ``dual emitters'' which are detected with both the \ha\ and \oiii\ lines. 
The numbers of galaxies in each sample are summarized in Table \ref{sec:numbers}.

Figure \ref{numdist3} shows the number distribution of the same physical quantities 
as in Figure \ref{numdist} but now for the three subsamples.  
We perform a KS-test, and the resulting $p$-values are listed in Table \ref{pvalue1}. 
For almost all of them, except for only two particular cases, the $p$-values are greater than 0.05, 
and we can statistically consider that all the emitter samples are drawn from intrinsically
similar distributions.

\subsubsection{Two exceptions: High ${\rm sSFR_{UV}}$ of the dual emitters}
\label{exception}

The two exceptions are the comparison of ${\rm sSFR_{UV}}$ between 
the \ha-single-emitters and the dual emitters,  
and between the \oiii-single-emitters and the dual emitters. 
Figure \ref{numdist3} shows that the dual emitters 
tend to have slightly higher ${\rm sSFR_{UV}}$ 
as compared to the other two subsamples.

Considering the comparison between the \ha-single-emitters and the dual emitters, 
this result indicates that the star-forming galaxies with relatively stronger \oiii\ emission lines 
tend to have higher star formation activity with respect to their stellar masses. 
This can be understood by the following arguments. 
Since high ${\rm sSFR_{UV}}$ produces more UV flux per volume, 
it leads to more extreme ISM condition characterized by the higher ionization parameter 
and thus showing the strong \oiii\ emission line with respect to \ha\ \citep{kewley15}. 
Therefore, the dual emitters, which tend to have the stronger \oiii\ emission lines than the \ha-single-emitters,  
are biased towards higher ${\rm sSFR_{UV}}$.

On the other hand, 
a possible difference between the \oiii-single-emitters and the dual emitters is not straightforward to interpret 
because the \oiii-single-emitters should also have the stronger \oiii\ emission line with respect to the \ha\ emission line. 
At lower ${\rm sSFR_{UV}}$ regime ($\lesssim$$-9.0$\ ${\rm [yr^{-1}]}$), 
the \oiii-single-emitters show a larger fraction of galaxies as compared to the dual emitters. 
This may be caused by a contribution of faint AGNs, 
but in order to investigate the presence of AGNs in our sample,
deep spectroscopy and line diagnostic analysis would be necessary.
We here also note that 
there is a contamination of the \oiii$\lambda$4959 and \hb\ emitters 
in our \oiii-single-emitters as mentioned in \S \ref{NBHcontamination}. 
These emitters would be misclassified as the \oiii-single-emitters  
even if they actually have the strong \ha\ emission line. 
In such a case, 
the stellar masses might be overestimated 
because of the contribution of \ha\ (and \nii) fluxes to the $K$-band fluxes. 
Therefore, the \oiii$\lambda$4959 and \hb\ emitters 
could contribute to the lower sSFRs seen in the \oiii-single-emitters.

At this point, we are not able to further investigate 
the cause of such difference, and we leave it for future investigation.
However, it should be stressed that except for the two particular cases, the physical properties of
the three emitter subsamples 
are not statistically different.

\subsection{Biases due to the NB selection}
\label{possiblebias}

Finally, we discuss possible biases introduced by the NB selection, 
which are based on the flux and EW of emission lines.

As shown in \citet{holden14} and \citet{shimakawa15}, 
the \oiii/\hb\ (\oiii/\ha) ratio is correlated with sSFR of star-forming galaxies 
in the sense that the galaxies with higher sSFRs tend to show the larger \oiii/\ha\ ratios. 
Given the fact that the EW of \ha\ is directly proportional to sSFR (e.g. \citealt{leitherer99}),  
the flux- and EW-limited sample might be biased towards the galaxies 
with the larger \oiii/\ha\ ratios.   
In such a case, the \ha\ and \oiii\ emitters might show similar properties 
because the \ha\ emitters tend to consist of the galaxies with the relatively strong \oiii\ emission line.

\citet{sobral14} investigated the relation between the rest-frame EW (\ha+\nii) and stellar mass 
for the \ha\ emitters at $z$=0.4, 0.8, 1.5, and 2.2 obtained by the HiZELS project.  
 They found that the \ha\ emitters at $z$$\sim$1--2.2 
 distribute well above the EW (\ha+\nii) cut of 25 \AA\ 
 up to a stellar mass of log(${\rm M_*/M_{\odot}}$)$\sim$11.5 
 (Figure 3 in \citealt{sobral14}).  
The relation between the rest-frame EW and stellar mass 
was also investigated for the \oiii+\hb\ emitters by \citet{khostovan16}. 
Their results show that the \oiii+\hb\ emitters at $z$$>$1 
have much higher EW than the EW selection limit.    
 Therefore, it is expected, in the first place, 
 that our samples are not strongly affected by the EW-cut.  
 Moreover, we show that the \oiii/\ha\ ratios of our samples 
 are roughly consistent with those of star-forming galaxies at the same epoch in the literature 
as in \S \ref{compareflux}. 
 The \oiii/\ha\ ratios do not seem to be largely different among the samples of star-forming galaxies selected by different methods. 
 This indicates that our NB-selected samples are not necessarily biased towards galaxies with higher \oiii/\ha\ ratios.

\section{Conclusions}
\label{summary}

We use the NB-selected galaxy catalog at $z$=2.23 obtained by the HiZELS project, 
and construct the two galaxy samples of \ha\ and \oiii\ emission line galaxies
by applying the same line flux limit. 
We derive the global physical properties of these emitters,  
and compare the number distribution of a stellar mass, dust extinction ($A_{\rm FUV}$), ${\rm SFR_{UV}}$, and ${\rm sSFR_{UV}}$ 
between the two samples.
The resulting $p$-values from a KS-test indicates that the \ha\ and \oiii\ emitters 
are drawn from the same parent population.
The two galaxy populations cover almost the same ranges of the integrated properties at $z$$\sim$2. 

We also divide the whole sample into three subsamples, namely,
the galaxies detected with either \ha\ or \oiii\ alone, and the galaxies detected with both lines. 
Again, a KS-test does not show any significant differences among the three subsamples, 
except for the dual emitters, which tend to be biased to higher ${\rm sSFR_{UV}}$ as compared to the other two subsamples. 
It is indicated that the strong \oiii\ emission lines are likely to be related to high star formation activities
(and thus high ionization parameters)  
of star-forming galaxies at $z$$\sim$2. 
Note, however, 
the \oiii\ and \ha\ emitters used in this study could harbor low luminosity AGNs, 
especially the \oiii-single-emitters with low sSFR as discussed in \S \ref{exception}, 
and spectroscopic observations are necessary to confirm the presence of AGNs.

In summary, the \oiii\ emitters trace almost the same galaxy populations as the \ha\ emitters
at $z$$\sim$2, and therefore we argue that the \oiii\ emission line can be used as an  
indicator of normal star-forming galaxies at high redshifts. 
Our results support the importance and the effectiveness of \oiii\ emitter surveys 
at $z$$\gtrsim$3, where \ha\ emission is no longer effectively observed from the ground.

\section*{Acknowledgements}
We thank the anonymous referee for 
careful reading and comments that improved the clarity of this paper.  
TLS acknowledges to Prof. Lisa Kewley 
for kindly accepting the visit  
to Mt. Stromlo Observatory for three months. 
The major part of this paper was written during the stay. 
TLS is also grateful to her group members for the many ways in which they helped.    
TK acknowledges the financial support in part by a Grant-in-Aid for the
Scientific Research (Nos.\ 21340045 and 24244015) by the Japanese Ministry of Education,
Culture, Sports and Science.
DS acknowledges financial support from the Netherlands Organisation for Scientific research (NWO) through 
a Veni fellowship, from FCT through a FCT Investigator Starting Grant and Start-up Grant (IF/01154/2012/CP0189/CT0010) 
and from FCT grant PEst-OE/FIS/UI2751/2014.
IRS acknowledges support from STFC (ST/L00075X/1), 
the ERC Advanced Grant DUSTYGAL (321334) and a Royal Society/Wolfson Merit Award.  
PNB is grateful for support from the UK STFC via grant ST/M001229/1.

\bsp	
\label{lastpage}
\end{document}